# INTEGRATION AND INTEROPERABILITY ACCESSING ELECTRONIC INFORMATION RESOURCES IN SCIENCE AND TECHNOLOGY:
*the proposal of Brazilian Digital Library*

Carlos Henrique Marcondes
*Universidade Federal Fluminense, Depto. de Ciência da Informação, R. Lara Vilela, 126, Niterói-RJ-Brazil*
*Email: marcondes@alternex.com.br*

Luis Fernando Sayão
*Centro de Informações Nucleares, R. General Severiano, 90, Rio de Janeiro-RJ-Brazil*
*Email: lsayao@cnen.gov.br*

Abstract: This paper describes technological and methodological options to achieve interoperability in accessing electronic information resources, available in Internet, in the scope of Brazilian Digital Library in Science and Technology Project - BDL, developed by Brazilian Institute for Scientific and Technical Information - IBICT. It stresses the impact of the Web in the publishing and communication processes in science and technology and also in the information systems and libraries. The work points out the two major objectives of the BDL Project: facilitates electronic publishing of different full text materials such as theses, journal articles, conference papers, "grey" literature - by Brazilian scientific community, so amplifying their nationally and internationally visibility; and achieving, through a unified gateway, thus avoiding a user to navigate and query across different information resources individually. The work explains technological options and standards that will assure interoperability in this context.

## 1. INTRODUCTION

"A research laboratory without a library is like an animal without brain: physical activities keep functioning but there is lack of memory coordination and will" Zilman (1979, p 115).

The convergence and integrated use of technologies for communication, computing and electronic publishing, for which Internet is a paradigm, have contributed, in recent years, to create a new environment for knowledge accessing, exchange, cooperation and promotion in global scale. We are in the middle of a process which consequences we cannot yet evaluate completely.

New knowledge media, completely different from printed material, are developed every day.

From the point of view of information as the basis to academic activities in S & T, Internet brings facilities that overcomes the concept of bibliographic information, based on printed documents such as journals articles, conference papers, theses and dissertations. New electronic information resources are now available to academic community besides the traditional paper resources, as multimedia documents, discussion lists, electronic newsgroups, online conferences, images (from satellites, microscopes, in real time), motion picture models, "eprints" archives, etc. These resources are both research and communication facilities thus enabling electronic publishing and immediate communication of research results.

More than just information resources, these new resources available through Internet are mainly "knowledge tools" according to Pierre Lévy (1993) sense. They open new intellectual and cognitive perspectives that overcomes tremendously the one offered by paper documents. For many authors Internet represents a breakage of paradigms comparable to the invention of press by Gutemberg. This change affects science communication too. Internet is a multidirectional and interactive communication device reaching the whole world: everyone can publish, what is published is immediately available, an author can receive comments about the information published from everywhere. A scholar author wishes maximum diffusion to his/her publications. To be cited by other authors is a measure of the impact of his/her research. Recent surveys confirm that electronic publishing is much more cited that publishing in paper: *"The mean number of citations to offline articles is 2.74, and the mean number of citations of online articles is 7.03, an increase of 157% (Lawrence)*. To develop mechanism to facilitate the electronic publishing for the Brazilian scholars, in order to enhance its visibility, is, therefore a fundamental aspect for the development and the maturity of the Brazilian research effort.

So, the paradigms of scientific communication, based on printed scientific periodicals with a schema of peer-review and the monopoly of great scientific publishers, are now suffering impacts with the raise of Internet and has been questioned by the scientific community worldwide. Since the emergence of the first electronic archive of preprints, or eprints, the ArXiv, in Los Alamos National Laboratory, created in 1991 by physicist Paul Ginsparg (Ginsparg, 1991), the international scientific community offers a practical alternative to publish their works. A list that shows the dimensions and amplitude of electronic archives worldwide can be founded in http://www.osti.gov/eprints/ppnbrowse.html.

Electronic archives have been articulated through the OpenArchives Initiative (http://www.openarchivesinitiative.org).

All this changes are deeply influencing the conception and the functioning of automated information systems, especially those that focus on research in S&T. In the last decade information develop made possible the establishment of a new stage to those systems: from isolated systems based on bibliographic records pointing to paper documents, those systems nowadays turn to distributed retrieval of digital objects – full text, motion picture, sound – establishing a new paradigm for publishing, directly in Internet, and achieving interoperability between heterogeneous and global distributed information resources.

With the Brazilian Digital Library Project (BDL), the Brazilian Institute for Scientific and Technical Information ( IBICT) intends to open the possibility to promote and to provide means to the Brazilian scientific community to publish routinely directly in Internet, thus enhancing its visibility nationally a internationally, optimizing scientific communication flow and reducing knowledge generation cycle.

On the other way, Brazil has accumulated experiences that are expressive, despite still isolated, in creating digital libraries and information repositories directly in the web. Some examples are Prossiga – Research Information and Communication Program - (http://www.prossiga.br), Scielo – Scientific Electronic Library Online - (http://www.scielo.br ), USPs – University of São Paulo - theses repository and e-archive at IMPA (Pure and Applied Mathematics Institute). As soon as Brazilian experiences multiply, interoperability becomes a critical problem: how to cross search all these resources with a minimum effort by a user, from a single gateway?

## 2. QUESTIONS OF INTEROPERABILITY

A problematic aspect of our culture related to information is the so-called phenomenon of "information overload", the enormous quantity of information produced and available from different social activities, making hard its identification, access and use. In the emergence of the so called information society, the value of information as a commodity for any activity, be it an economic decision making, a cultural or education process, a scientific or technological research, is directly concerned with its potential of economically guiding the expenditure of energy to perform it. To achieve all this potential, information relevant to a problem solve must be available at the right time. It is no use if those who need information do not know its existence or do not succeed in finding it.

This situation assumed critical dimensions with the raise of Internet. A notice published in EDUPAGE Bulletin in Portuguese in April, 4, 1998, stresses the problem of information discovery in Internet and comments the results of a survey on search engines performance:

¨FINDING A NEEDLE IN THE WEB

*A survey performed by NEC Research Institute assures that Internet explodes to more than 320 million Web pages, an estimate that does not includes*

*millions of pages with access restricted by passwords or firewalls that restricts the access to browsers or search engines. The survey shows that HotBot search engine has the most comprehensive index of the Web, but covers only 34 per cent of indexed pages. The coverage of some of the other search engines includes: AltaVista (28%); Northern Light (20%); Excite (14%); Lycos (3%)."*

A novelty about search engines, which seems promising are GOOGLE and CLEVER projects (http://www.google.com), with their proposal of ranking search results based number of links from other pages to this page (Clever, 1999).

The enormous quantity of information stored and available through Internet makes information discovery a crucial problem. Different strategies to face the information overload problem brought by Internet can be outlined nowadays as general purpose search engines (AltaVista, Excite, Lycos, Infoseek, Yahoo, etc), specialized information locators services as GILS or subject gateways as American SIGNPOST (http://www.signpost.org), British OMNI (http://www.omni.ac.uk) and SOSIG (http://ww.sosig.ac.uk) and Brazilian Prossiga (http://ww.prossiga.br) and Lis – Heath Information Locator - (http://ww.bireme.br).

All alternatives, general purpose search engines, specialized information locators services or subject gateways offers partial and limited solutions to information resource location in Internet, mainly those of interest to science and technology. Search engines deficiencies are well known and discussed in literature (Sneiderman, 1997): low quality from automatic indexing, resulting in high retrieval of information but a low rate of precision; covering only part of Internet; search engines are not specialized; they index isolated HTML pages, not entire resources; besides this, a great amount of information available in Internet are stored as registers in data bases, hidden from search engines robots by interfaces that demands user interaction.

Isolated digital libraries and subject gateways in turn are only partial solutions to the problem of accessing information resources of interest to science and technology published in Internet: they are limited to their own holdings. Besides great investments have been made in describing, indexing, storing and make available great amount of information through digital libraries and subject gateways, more and more information resources are created and made available continuously, stored in isolated servers, operated by different search interfaces, restricted to a user to perform a tedious search, site by site, to find relevant information.

From an academic user or a researcher point of view, an interesting aspect would be to submit his/her information needs to a single interface and receives information from different sources, in a consolidated way. This question has been discussed more and more in literature: "digital libraries federation" and "distributed archives" (Liu, 2001), "confederated digital libraries" (Leiner, 1998), "distributed subject gateways" (IMesh, 1999), "networked digital library" (Davis, 1995), "multiple information sources" (Paepcke, 2000) "cross-searching", "heterogeneous distributed databases", "metaserches" (Gravano, 1996), etc.

The problem is first posed in "An Intenational research agenda for digital libraries", of 1998 - a join research agenda of NSF (USA) and European Union – in three thematic work groups: "global resource discovery", "interoperability" and "metadata". Nowadays the problem is addressed by different research groups, as Joint NSF – JISC International Digital Libraries Research Programme, Imesh consortium (http://www.desire.org/html/subjectgateways/community/imesh/), the Scout Project, and for practical iniciatives as OpenArchives Initiative, Arc - Cross Archive Searching Service, NCSTRL (Cornell University, USA), NDLTD (Virginia Tech University, USA), Digital Library Federation (a consortium of American digital libraries), ROADS (UKOLN JISC, England). The theme emerges in literature under different forms but all of them converge to concepts of integration and interoperability in digital libraries, which means the *possibility of a user search different and heterogeneous information resources, stored in different servers in Web from a single interface, without taking notice of where and how theses resources are stored.*

Today there are today different alternatives of interoperability and integrated access to heterogeneous information resources available in the Web. These alternatives can be basically grouped in two, although there isn't a standard designation widely accept: the user interfaces distribute a search to different servers and the user interface searches a centralized metadata database. In both alternatives the user interacts only with a single Web interface where his/her search is submitted.

In the first alternative, the interface distributes the search to different servers (named also broadcast search) according to a standard protocol; these servers are identified by the interface as capable to answer satisfactorily to the search and the results are consolidated and integrated. An example of this alternative is the well known Z39.50 protocol, used to provide interoperability among automated library catalogs. This alternative shows the followings advantages: new data providers can be added, provided that the data providers adhere to the standards adopted, with a minimum need of reconfiguration. As disadvantages this alternative requires the server to run the protocol software to be searchable. Some of these protocol software are heavy resources consumers, as Z39.50 (Troll, 2001); in some cases there is need for dedicated servers, as the index servers that routes searches to servers capable to answer them. According to this alternative, some technological standards available are the following: Z39.50 (ISO/NISO), Whois++, LDAP, CIP, SDLIP, DIENST. This alternative is used in the following systems: NCSTRL (Cornell University, USA), NDLTD – Networked Digital Library of Theses and Dissertations - "federated search" (Powel, 1998), California Digital Library (http://www.cdlib.org/), Berkeley Environmental Digital Library, USA, ROADS, ISAAC/SCOUT Project (Stanford University, USA Z39.50 (ISO/NISO), Whois++, LDAP, CIP, SDLIP, DIENST. This alternative is used in the following systems: NCSTRL (Univ. Cornell, EUA), NDLTD – Networked Digital Library of Theses and Dissertations - "federated search" (Powel, 1998), California Digital Library (http://www.cdlib.org/), Berkeley Environmental Digital Library, EUA, ROADS, ISAAC/SCOUT Project (University of Stanford, EUA).

The second alternative, metadata of electronic documents are periodically gathered to a centralized searchable database. This is the well known schema of a union catalog. There alternative of sending metadata to a centralized database is largely known by Brazilian information community in systems/databases as LILACS/BIREME, SITE/IBICT and INIS/CIN. This schema presents the following advantages: new data providers can participate in the system provided that they adhere to the standards adopted; good performance due to searches submitted directly to service provider's local database. As disadvantages this schema presents: maintenance cost of common metadata database by service provider; need to synchronize data stored in data providers with data centralized in metadata database in service provider

The automatic gathering of metadata, named ¨harvesting¨ is more recent: metadata from various information providers became ¨visible¨ through standard protocols e are automatically collected periodically and stored in a data warehouse, where searches are submitted. Advantages of this schema are: new data providers can join to the system provided that they adhere to standards adopted, good performance due to searches submitted to local metadata database of service provider. Disadvantages are: maintenance by service provider of common metadata database, need to keep synchronized data stored in data providers with data harvested from them. Standards used in this option are: OpenArchives Harvest Protocol, Open Archives Metadata Set, Dublin Core, XML. Some outstanding experiences are: MARIAN digital library system (Virginia Tech. University), the NDLTD portal, OpenArchives Initiative, that through OAI harvest protocol permits metadata harvesting from compliant eprint archive servers; Arc – Cross Archive Searching Service, first service to provide integrated access to different electronic archives (http://www.arc.cs.odu.edu).

## 3. BRAZILIAN DIGITAL LIBRARY MODEL OF INTEROPERABILITY

The implementation of the BDB Project relies on a strong political action of integration, by IBICT, directed to the Brazilian most important content providers and science and technology information services. This integration will be accomplished by the priority project of Brazilian Digital Library: electronic publishing of full text in science and technology by Brazilian academic community and interoperability between different Brazilian information systems and services through a unified gateway, assuring as a fundamental point, the independence and autonomy of each member: metadata set, system/services configuration, standards must be the most simpler and the least onerous to different data providers, assuring their maximum organizational and technological independence.

Besides the accelerated development of Web, communication and information technologies, the project is built around consolidated, stable and trustable technologies as have been showed in various experiences similar to Brazilian Digital Library all over the world. Besides these requirements, theses technologies must be installed, maintained and, as necessary, modified, by IBICT information technology staff. These technologies must also be easily passed to project partners. This option has the objective to disseminate theses standards and protocols in Brazil so they can be adopted by future partners of BDL and by other information systems that intend to join different international systems and nets.

Other relevant policy of BDL project is the recommendation that its partners must always use of free or public domain software to web publication. There are different free software of quality that have been adopted by information systems and institutions all over the world.

BDL's model of interoperability is very like those of NDLTD portal (Suleman, 2001) and of Arc – Cross Archive Searching Service (http://www.arc.cs.odu.edu). Both systems perform metadata harvesting from data providers, maintaining a centralized metadata database. The BDL Internet gateway will materialize the Brazilian Digital Library in science and technology. The gateway will enable a researcher, through various interoperability mechanisms, integrated and unified access to different Brazilian heterogeneous information resources relevant to science and technology, without the need to search each information resource one by one.

In its first version, the BDL will encompass Brazilian electronic journals from Scielo gateway, electronic proceedings from Brazilian events in science and technology, electronic theses and dissertations from USP – University of São Paulo -, UNICAMP – University of Campinas, UFSC – Federal University of Santa Catarina -, PUC-Rio – Pontiffic Catholic University of Rio de Janeiro -, ENS/FIOCRUZ – Public Health National Scholl - and different Brazilian electronic archives.

The OpenArchives Initiative identifies two roles in an electronic publishing environment: data providers and service providers. These definitions will be used here to explain the BDL interoperability model; according to the OpenArchives Initiative:

**Data providers:** this definition would encompass all Brazilian information institutions having a Web site to provide electronic documents in science and technology. The site must include an electronic document submission environment too, where a author can fill a form with metadata about his/her document and can upload and store it. The site provides also facilities for searching for the metadata of documents in its collection. Sites of electronic archives complaint with OpenArchives Harvesting Protocol can also made available metadata of their collections to harvesting processes from different service providers. Typical examples of data providers would be CogPrints (http://cogprints.soton.ac.uk/), or all information institutions from who BDL would harvest metadata to a central metadata database to provide integrated access, as Scielo or Brazilian electronic archives as IMPA´s (http://www.preprint.impa.br/indexEngl.html)

**Service providers:** institutions providing added value services based on electronic documents available in different data providers. Value added services would be building of qualified databases from various sources, peer-review of electronic documents published in electronic archives, cross linking different sets of electronic documents. Typical examples are Arc (http://arc.cs.odu.edu/) which provides unified access to various electronic archives and what would be the BDL.

Due to its integrating role between different Brazilian systems and data providers, BDL´s interoperability model is based in two elements: a gateway with a single interface to submit queries and a metadata set to describe electronic documents and resources, providing an unified view of different documents sets.

The BDL gateway submission queries mechanism must encompass the main technological alternatives discussed previously: distributed search and search to a centralized metadata database built from metadata harvesting from Brazilian data providers. BDL gateway interfaces will incorporate a Z39.50 client providing integrated access through distributing queries to different library having a Z39.50 server. Any data providers having a Z39.50 server could be accessed from BDL gateway too. This interoperability option is showed in the following figure:

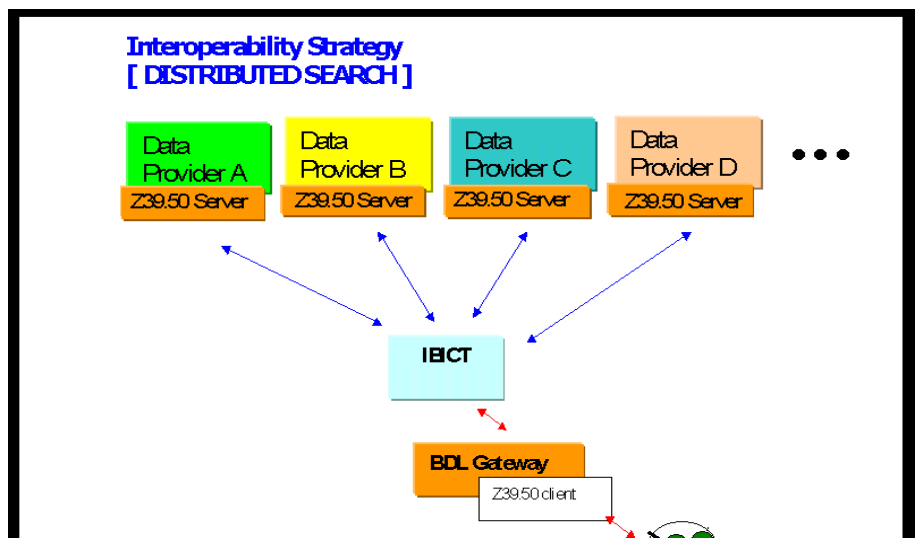

Besides interoperability through distributed searches via Z39.50 protocol, the BDL will maintain its own metadata database with metadata harvested from data providers not having a Z39.50 server. Harvesting process will be performed periodically, through OAI harvesting protocol. Other particular solutions are foreseen with the objective of do not burden data providers that are not still compliant with OAI harvesting protocol. These particular solutions include metadata gathering through FTP of HTML and text files.

Metadata gathered is stored in a common metadata database maintained in BDL site. This metadata database would be served by a Z39.50 server, in a model similar to other remote data providers databases, accessed through this protocol too. This second interoperability option will enable different preprints archives be installed in university departments, research institutions, scientific societies, electronic journals as well as cooperate projects as the Genome Project will be able to have their metadata collected by BDL. This option is illustrated in the following figure:

Other main element of BDL interoperability schema is its metadata set. This metadata set must mediate and integrate in an unified description different electronic documents types. The emerging standard relative to metadata is Dublin Core Element Set. It is the result of an intense international discussion and standardization process, maintained by active groups and forum named Dublin Core Metadata Initiative, having realized various meetings. Dublin Core is now used in different systems including OpenArchives Initiative.

Dublin Core metadata set includes thirteen descriptive elements (Dublin Core, 1999). Qualifiers to specify the meaning of elements are also supported (Dublin Core, 2000). Dublin Core metadata elements can be codified in formats as HTML and XML (Cox, 2000), (Beckett, 2000), (Beckett. 2001). As stressed in the Dublin Core proposal, metadata set must be so simple and intuitive in order to allow the authors of electronic document to describe it by themselves. Electronic documents submission environments operate in the same way: authors fills a form describing their documents before submitting a copy to an electronic archive.

BDL will use Dublin Core metadata sets, expanding its qualifiers to support specific characteristics of some types of documents, as electronic theses and dissertation, research reports, journal articles, conference papers.

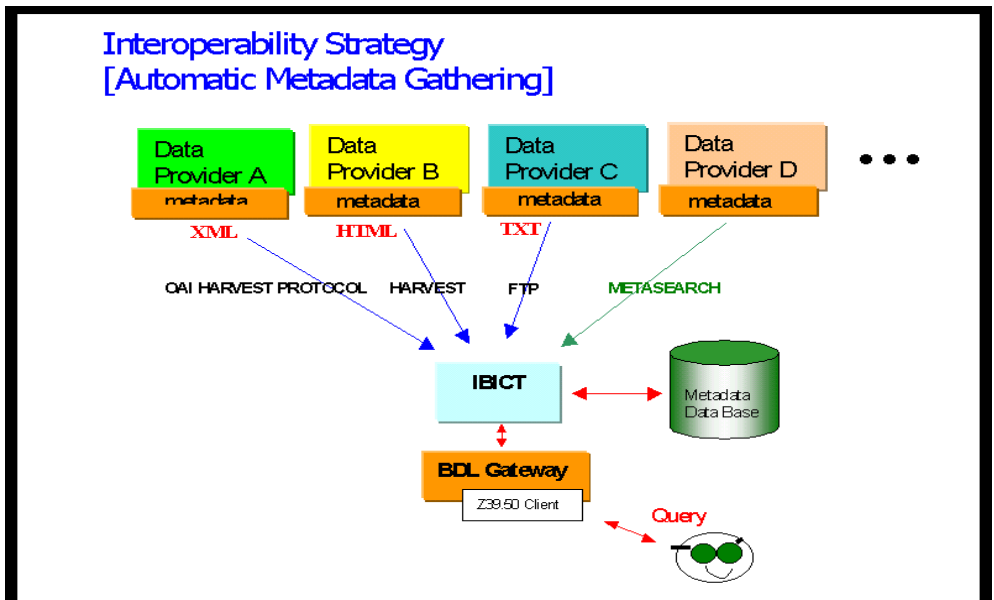

Figure 2 - Interoperability Strategy – Automatic Metadata Harvesting

# 4. CONCLUSIONS

International experiences in electronic publishing in the Web and interoperability between digital libraries are recent, some of them are now in development. Solutions to technical problems related to these questions are focus of research by information and computer science researchers. In spite of this research activity, it is evident that there are right now consolidated standards an technologies that can constitute strong bases to build operational systems. The interoperability questions addressed by BDL project are still restricted to a technological focus: interoperability between heterogeneous information resources in Web have different dimensions – semantic, politics/human, intercommunity, international, interlinguistic (Powel, 1998) -, as stresses Miller.

The success of the BDL projects, in short or long terms depend on IBICT institutional attitude regarding monitoring research, technological and standard development. This wide framework is too complex for IBICT alone. To face these questions IBICT must take over an articulation role between different Brazilian information system partners and academic community, around join activities and a research agenda including main questions related to BDL project, as suggested in the project:

> *"Monitoring and articulation with international forums regarding questions as metadata, interoperability, electronic publishing, copyright of electronic publishing, cross linking, etc, as World Wide Web Consortium, Dublin Core Metadata Initiative, Open Archives Initiative, Digtal Library Federation. With the aim of monitoring the development of digital libraries technologies and the discussion around this question, it is necessary to approach to organizations or networks that operate networked or heterogeneous digital libraries. It is also important too the participation of IBICT to international forums where the discussion relative to these questions take place. So, it is recommended:*
>
> *- To suggest a research agenda and a grant found for research themes suggested above to Brazilian scientific community, including universities, research institutions and partners of BDL project*
>
> *- To suggest a national forum on digital libraries, to promote discussion and exchange of experiences*
> *- To suggest a educational program"*
> (IBICT, 2001, p. 16).

The BDL project, besides its commitment with new information services to Brazilian academic community, raises also some problems regarding to the planning of information in science and technology. Since 80's, with Programmed Action in Science and Technology Information (Brasil, 1984) and with PADCTs – Science and Technology Development Support Program - (Brasil, 1985), there were no more comprehensive STI planning documents. In this context is important point out that neither the Green Book of the Brazilian Information Society Program (2000) nor the Green Book of the Science, Technology and Innovation Program (2001) includes questions regarding to STI.

At present the scientific communication stands more and more on information technology. BDL project brings questions that would constitute a framework to build an integrated information environment. Its is clear IBICT's role to plan and promote this technological infrastructure. This environment would provide to users overall facilities to electronic publishing and wide access to electronic resources. Components of this technological infrastructure would be: electronic publishing and storage facilities, cross linking of electronic documents, persistent electronic addresses, authority and thematic description databases, metadata systems, information discovery facilities and long term electronic documents preservation mechanisms.

An comprehensive initiative as BDL must be managed by a steering committee composed by representatives of strategic Brazilian information systems partners and academic community to whom BDL is addressed. This steering committee must meet periodically, approve past activities report and develop a work plant do BDL.

Although ambitious, BDL objectives are all feasible. The complexity of the project is more organizational one than technological. BDL project can start with a low investment due to technologies and standards already developed internationally. BDL emphasis in electronic publishing methodologies and interoperability between heterogeneous information resources are necessary to provide to Brazilian academic community an infrastructure compatible with international standards. This will enable Brazilian academic community be inserted in the international information flux and provide an increase visibility to Brazilian science. Science and technology information systems and among then digital libraries like BDL, must help authors of documents to improve their visibility, adopting technologies that maximize broad integration and interoperability between information systems.